# Modelling of standard and specialty fibre-based systems using finite element methods

Natascia Castagna*[a], Jacques Morel[a], Luc Testa[b], Sven Burger[c,d]
[a]Federal Institute of Metrology METAS, Lindenweg 50, 3003 Bern-Wabern, Switzerland; [b]Ecole Polytechnique Fédérale de Lausanne, EPFL PH D2 375 Station 3, 1015 Lausanne, Switzerland; [c]JCMwave GmbH, Bolivarallee 22, 14050 Berlin, Germany; [d]Zuse Institute Berlin, Takustraße 7, 14195 Berlin, Germany

*natascia.castagna@metas.ch; phone 0041 583870643; www.metas.ch

## ABSTRACT

We report on the investigation of an approach for modelling light transmission through systems consisting of several jointed optical fibres, in which the analytical modelling of the waveguides was replaced by Finite Element Modelling (FEM) simulations. To validate this approach we first performed FEM analysis of standard fibres and used this to evaluate the coupling efficiency between two singlemode fibres under different conditions. The results of these simulations were successfully compared with those obtained using classical analytical approaches, by demonstrating a maximum loss deviation of about 0.4 %. Further, we performed other more complex simulations that we compared again to the analytical models. FEM simulations allow addressing any type of guiding structure, without limitations on the complexity of the geometrical waveguide cross section and involved materials. We propose as example of application the modelling of the light transmitted through a system made of a hollow core photonic crystal fibre spliced between two singlemode standard optical fibres, and qualitatively compare the results of the simulation with experimental results.

**Keywords:** Finite element modelling, modal distribution, specialty fibres

## 1. INTRODUCTION

Important quantities like the insertion loss between jointed multimode fibres or the bandwidth depend on how the different guided modes are populated and propagate all along the fibre. This strongly depends on how the light source is coupled into the fibre, and on how the modal selection occurs at the interface between different sections of concatenated fibres. A modelling of these effects is usually performed using analytical approaches to quantify the modal distribution and then to calculate the overlap integrals. These techniques are very well applicable when considering simple structures, but may become impractical when dealing with more complex waveguides, like micro-structured fibres. In this work we investigate a different approach, in which the analytical modelling of the waveguides was replaced by FEM simulations, allowing addressing any type of guiding structure, without limitation in its complexity. To validate this approach, we performed different FEM analysis of standard fibres and the results of these simulations were successfully compared with those obtained using classical analytical approaches. First, we evaluated the coupling efficiency between two singlemode fibres under different conditions, namely the misalignment of the fibre optical axis with respect to the direction of propagation of the light source; the results are presented in Section 2. Then, in light of the results obtained with this first test, we performed a more complex simulation, for which we were inspired by an analytical work published by Mafi *et al.* in 2011[1]. In their work the authors make use of a multimode fibre as connector between two singlemode fibres of different mode-field diameters to optimize the mode matching and thus to reduce the coupling losses; we present in Section 3 the comparison between their analytical models and our FEM simulations. Finally in Section 4 we propose, as example of application, the modelling of the light transmitted through a system made of a hollow core photonic crystal fibre spliced between two singlemode standard optical fibres, and qualitatively compare the results of the simulation with experimental results.

## 2. COUPLING EFFICIENCY BETWEEN TWO SINGLEMODE FIBRES

The aim of this simulation was to model the coupling efficiency between two singlemode fibres. We considered two singlemode fibres, whose optical axis were misaligned, by applying either an angular, or a lateral offset. The two cases are schematised in Figure 1.

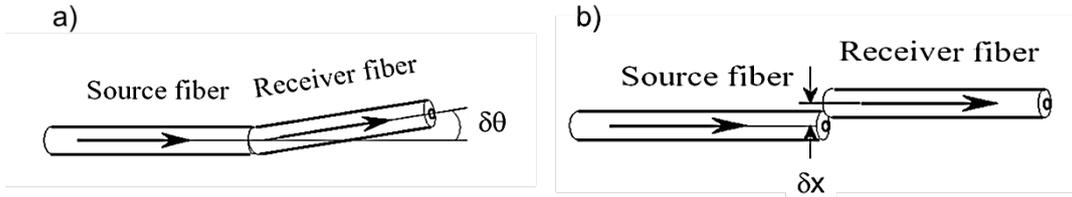

Figure 1. Angular (a) and lateral (b) misalignments between two fibres.

The FEM software package JCMsuite[2] was used to model the fibres, to calculate the guided eigenmodes as well as the overlap integral between the two fibres. The parameters used in the simulation were as follows:

| | |
|---|---|
| Core diameter $\phi_c$ / µm | 9 |
| Cladding diameter $\phi_{cl}$ / µm | 120 |
| Core refractive index $n_c$ | 1.452 |
| Cladding refractive index $n_{cl}$ | 1.443 |
| Wavelength $\lambda$ / nm | 1550 |

### 2.1 Angular misalignment

We considered in this first example an angular misalignment between the two optical fibre axis, see Figure 1a). The result of the simulation is shown in Figure 2, together with the result obtained with an analytical approach for the same configuration and parameters. The analytical model was developed by RP Photonics[3]. The vertical axis shows the normalised coupling efficiency.

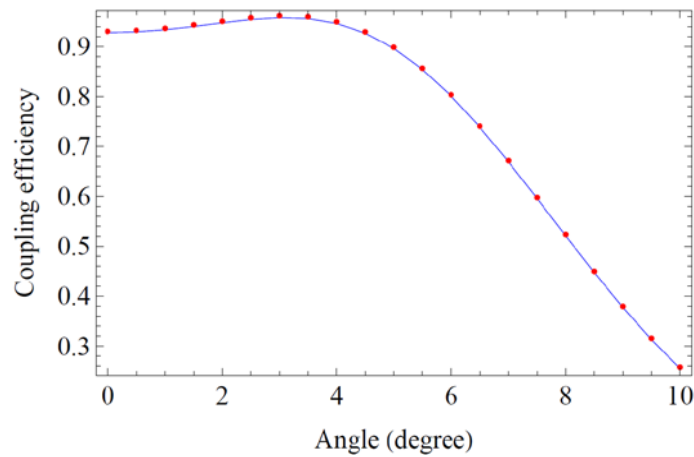

Figure 2. Coupling efficiency between two identical singlemode fibres with an angular misalignment between the two optical axes. In blue the result of the FEM modelling and in red the result of an analytical approach.

The agreement between the two approaches is high, with a maximum deviation of about 0.4 %.

## 2.2 Lateral misalignment

As a second example, the influence of a lateral misalignment of the two optical axes along the *x*-axis to the coupling efficiency was investigated (see Figure 1b). Once again the FEM simulations were compared with the results of an analytical approach and are shown in In Figure 3.

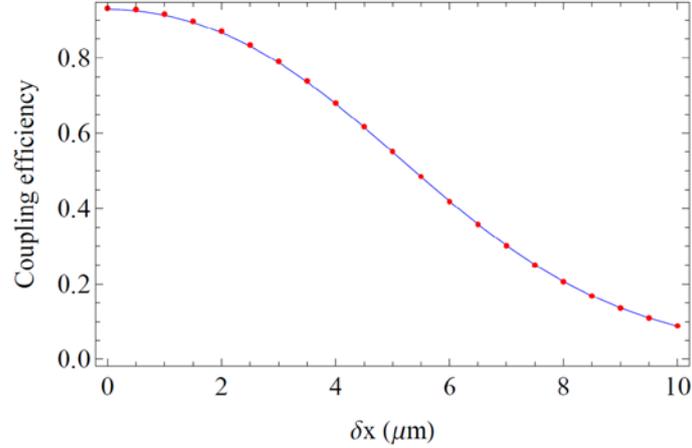

Figure 3. Coupling efficiency between two identical singlemode fibres with a lateral misalignment along the x-axis. In blue, the results of our FEM modelling and in red the results of an analytical approach.

The agreement between the two evaluation methods is better than 0.4%.

## 3. LOW-LOSS COUPLING BETWEEN FIBRES: ANALYTICAL VS FEM ANALYSIS

In light of these results, we investigated the performances of a more complex system, inspired by the analytical work published by Mafi *et al.*[1]. The aim of that work was to demonstrate that the coupling losses between two singlemode fibres with different mode-field diameters may be reduced by inserting between them a piece of gradient-index multimode fibre having a well adapted length (see Figure 4). They developed for that purpose an analytical model, which allowed determining the coupling efficiency and its dependency with the length of the multimode fibre.

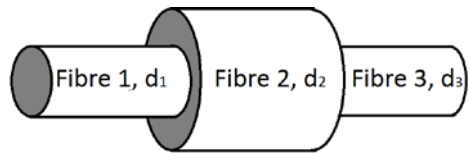

Figure 4. Scheme of the optical system used to reduce the coupling losses between two singlemode fibres (Fibre 1 and Fibre 3) with dissimilar mode-field diameters. Fibre 2 is a multimode fibre, which acts as a mode filed diameter adapter. $d_1$, $d_2$ and $d_3$ are the respective core diameters of the three fibres.

The coupling efficiency oscillates, according to the cited paper, with a period length $Z$ given by:

$$Z = \frac{\pi \cdot d_2/2}{\sqrt{2 \cdot \Delta}} \quad , \tag{1}$$

where $\Delta$ is the index step of the gradient index fibre defined by:

$$\Delta = \frac{n_c^2 - n_{cl}^2}{n_c^2} . \tag{2}$$

Considering the same parameters as those used by Mafi et al.[1], namely $d_1 = d_3 = 8.2$ μm, $d_2 = 50$ μm, and taking $n_c=1.452$ and $n_{cl}=1.443$ for the index of refractions of the core and of the cladding of fibres 1 and 2, a period length of $Z = 0.7$ mm was calculated using Eq. (1). This same structure was modelled using the FEM software package JCMsuite[2] to calculate the eigenmodes of the multimode fibre and the transmission through the three fibres with the overlap integrals. Figure 5 shows the resulting transmission through the optical system as a function of the length of the multimode fibre. The periodical dependence of the transmission to the fibre length is clearly visible, whit a period length of $Z_{FEM} = 0.67$ mm, which is in a very good agreement with the analytical solution.

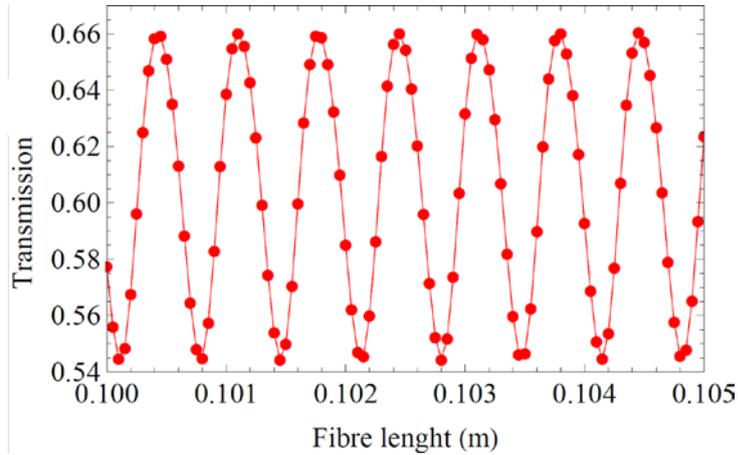

Figure 5. Transmission through the three fibres system shown in Figure 4 as a function of the length of the multimode fibre.

## 4. APPLICATION TO SPECIALTY FIBRES

We propose as example of application the modelling of light transmission through a setup including a photonic crystal fibre (PCF) which is spliced to a standard single mode fibre (SMF) on both sides, in the same setup as depicted in Fig. 4. We have performed experimental transmission measurements of this setup with the PCF filled with acetylene. Figure 6A shows a microscopic image of the PCF cross section before splicing, with a core diameter of about 7.3 μm. The SMF core diameter is 8.2 μm. Figure 6B shows the measured transmission as function of the frequency of the exciting laser. The wavelength is centred around 1542.4 nm and scanned in a range of about 3.5 GHz (few pm). The measured transmission signal exhibits an intensity modulation with a period of about 0.5 GHz superposed to the acetylene Doppler absorption. Our numerical setup relates to this example, however, it does not accurately model the experimental setup nor the presence of the gas inside the PCF, which would be beyond the scope of this demonstrational study. Figure 6C shows the FEM triangular mesh[4] discretizing the cross section of a hollow core PCF with a central core of 7 missing cladding pores, with a cladding with a periodicity length of 2.52 μm, a pore strut width of 80 nm, corner rounding radius of 390 nm and 7 rings of cladding pores. A slight asymmetry of the geometrical setup is assumed (ellipticity of 2%) which yields a small difference of the propagation constants of the two (nearly degenerate) fundamental modes. The computed intensity distribution of one of these is visualized in Fig. 6D. For the corresponding FEM computation, higher-order finite elements (p=3) are used[5]. In our computational setup, we first compute the modes of the involved fibres (Fibre 1, 2, 3), then we compute the overlap integrals (which is done in a postprocess, utilizing the higher-order FEM discretization of the field distributions). From application of the respective transfer matrices to the incoming fibre mode, we compute the total power transmission through the setup. The transmission spectrum for a corresponding frequency scan is shown in Fig. 6E. Please note, that here only the two fundamental modes contribute: in the numerical study, all investigated higher-order propagation modes of the structure exhibited damping losses at least one order of magnitude larger than the damping losses of the fundamental modes, we have therefore only considered the fundamental mode propagation in the results shown in Fig. 6E.

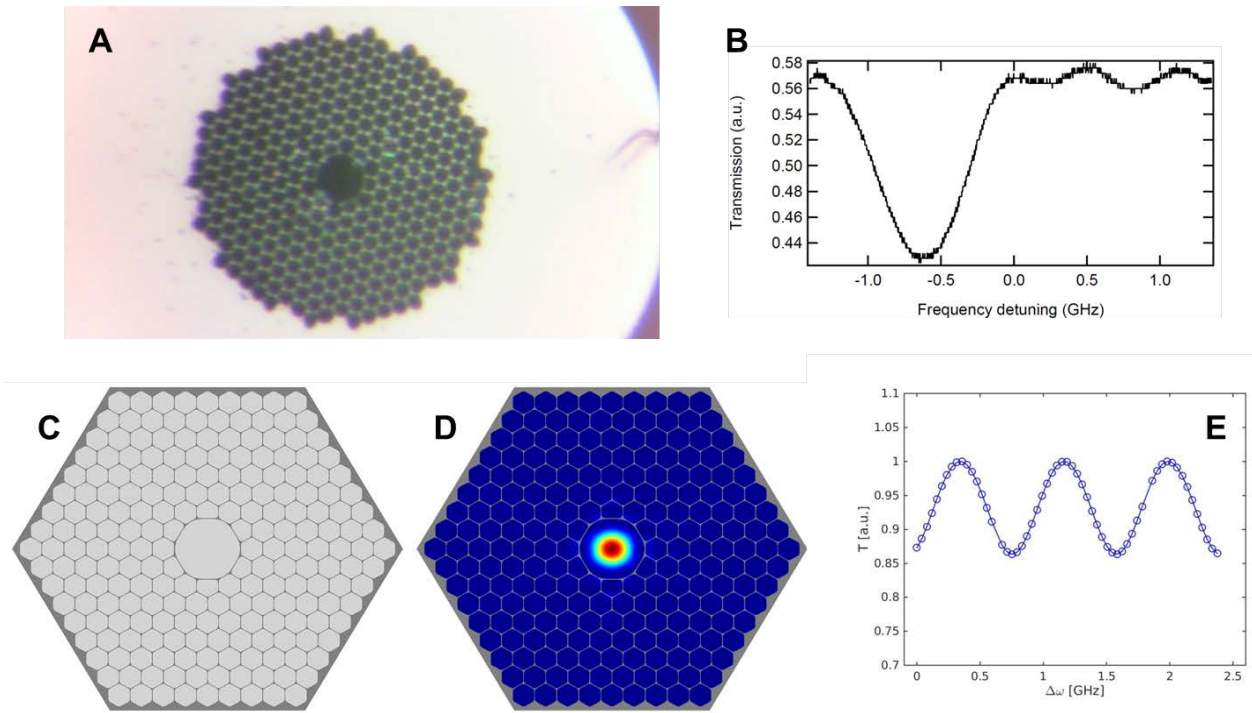

Figure 6. A: Microscopic image of the investigated PCF. B: Detector signal proportional to the transmission through the PCF shown in A, spliced in between two standard SMF and filled with acetylene. C: Geometrical model of the cross section of a HCPCF (light grey: air, dark grey: silica) with triangular mesh for the FEM discretization. D: Visualization of the electric field intensity distribution of a fundamental mode in the PCF (red: high intensity, blue: low intensity) with superimposed cladding structure for visualization purposes (grey). E: Simulated transmission through a SMF – HCPCF – SMF setup.

## 5. CONCLUSION

FEM modelling techniques can be advantageously used to evaluate the transmission properties of complex waveguide-based systems, especially in situations where the derivation of analytical solutions is getting too complicated, as it is the case when using specialty or photonic crystal fibres. Another interesting domain of application is related to the metrology of multimode waveguides, especially towards the definition of optimum templates for the modal distribution in multimode fibres, to ensure repeatable power and loss measurements.

## 6. ACKNOWLEDGMENTS

This project has received funding from the EMPIR programme co-financed by the Participating States and from the European Union's Horizon 2020 research and innovation programme under grant agreement number 14IND13 (PhotInd).